%% file: main_AJ_submit20250320.tex
\documentclass[twocolumn, times]{aastex631}
\usepackage{xspace}
\usepackage{url}
\usepackage{color}

\newcommand\CC{2001~CC$_{21}$\xspace}
\newcommand\KY{1998~KY$_{26}$\xspace}

\received{February 25, 2025}
\revised{March 20, 2025}
\submitjournal{AJ}

\shorttitle{Size Constraint on 1998~KY$_{26}$ via VLT/VISIR Non-detection}
\shortauthors{Beniyama et al.}

\begin{document}
\title{Size Constraint on Hayabusa2 Extended Mission Rendezvous Target 1998~KY$_{26}$ via VLT/VISIR Non-detection}

% Author contributions (by J.B.) ================================================
%   Jin Beniyama       : led the project
%   Thomas G. Muller   : prepared VLT/VISIR proposal,
%                        performed the TPMs, 
%                        contributed to the interpretation, 
%                        contributed to writing the manuscript
%   Marco Delbo        : prepared VLT/VISIR proposal, 
%                        provided the TPM code to J.B., 
%                        contributed to the interpretation, 
%                        contributed to writing the manuscript
%   Eric Pantin        : performed VLT/VISIR analysis, 
%                        contributed to writing the manuscript
%   Olivier R. Hainaut : prepared VLT/VISIR proposal, 
%                        contributed to writing the manuscript
%   Marco Micheli      : prepared VLT/VISIR proposal,
%                        contributed to writing the manuscript
%   Michael Marsset    : prepared VLT/VISIR proposal, 
%                        contributed to writing the manuscript
% Author contributions (by J.B.) ================================================

\correspondingauthor{Jin Beniyama}
\email{jbeniyama@oca.eu}

\author[0000-0003-4863-5577]{Jin Beniyama}
\affiliation{Université Côte d'Azur, 
    Observatoire de la Côte d'Azur, CNRS, Laboratoire Lagrange, Bd de l'Observatoire, 
    CS 34229, 06304 Nice Cedex 4, France
}
\affiliation{
    Institute of Astronomy, Graduate School of Science, 
    The University of Tokyo, 2-21-1 Osawa, Mitaka, Tokyo 181-0015, Japan
}

\author[0000-0002-0717-0462]{Thomas G. M{\"u}ller}
\affiliation{
    Max-Planck-InstitutfürextraterrestrischePhysik, Giessenbachstraße, Postfach1312, 85741 Garching, Germany
}

\author[0000-0002-8963-2404]{Marco Delbo}
\affiliation{Université Côte d'Azur, 
    Observatoire de la Côte d'Azur, CNRS, Laboratoire Lagrange, Bd de l'Observatoire, 
    CS 34229, 06304 Nice Cedex 4, France
}
\affiliation{
    School of Physics and Astronomy, University of Leicester, Leicester LE1 7RH, UK
}

\author[0000-0001-6472-2844]{Eric Pantin}
\affiliation{
    Université Paris-Saclay, Université Paris Cité, CEA, CNRS, AIM, 91191, Gif-sur-Yvette, France
}

\author[0000-0001-6952-9349]{Olivier R. Hainaut}
\affiliation{
    European Southern Observatory, Karl-Schwarzschild-Stra{\ss}e 2, 85748 Garching-bei-M\"unchen, Germany
}

\author[0000-0001-7895-8209]{Marco Micheli}
\affiliation{
    ESA NEO Coordination Centre, Planetary Defence Office, 
    Largo Galileo Galilei 1, I-00044 Frascati (RM), Italy
}

\author[0000-0001-8617-2425]{Micha{\"e}l Marsset}
\affiliation{
    European Southern Observatory, Alonso de Córdova 3107, Santiago, Chile
}

\begin{abstract}
1998~KY$_{26}$ is a tiny near-Earth asteroid ($H=26.1$) discovered in 1998.
It has been selected as the target of the Hayabusa2 extended mission, which will rendezvous with 1998~KY$_{26}$ in 2031.
However, one of the most basic physical properties, size, remains poorly constrained, posing potential challenges for spacecraft operations.    
We aimed at constraining the size of 1998~KY$_{26}$ by means of thermal infrared observations.
We performed thermal infrared observations of 1998~KY$_{26}$ using the
ESO Very Large Telescope/VISIR on three consecutive nights in May 2024.
After stacking all frames, we find no apparent detection of 1998~KY$_{26}$ on the resulting images.
The upper-limit flux density of 1998~KY$_{26}$ is derived as 2~mJy at 10.64~$\mu$m.
From this upper-limit flux density obtained via non-detection,
we conclude that the diameter of 1998~KY$_{26}$ is smaller than 17~m with thermophysical modeling.
This upper limit size is smaller than the radar-derived 30 ($\pm$ 10)\,m.
Our size constraint on 1998~KY$_{26}$ is essential for the operation of the Hayabusa2 spacecraft
during proximity operations using remote sensing instruments as well as a possible impact experiment using the remaining projectile.
\end{abstract}


\section{Introduction} \label{sec:intro}
% 1. Hayabusa2 & extended mission
The Hayabusa2 spacecraft finished its main mission, the exploration of the near-Earth asteroid (NEA) (162173)~Ryugu in 2020, after
dedicated in situ remote-sensing measurements and experiments as well as a successful sample return \citep[e.g.,][]{Sugita2019, Watanabe2019, Kitazato2019}.
Currently, Hayabusa2 is in the cruise phase of its extended mission, 
which will consist in visiting two more NEAs. 
Recently, \KY has been selected as the rendezvous target of the Hayabusa2 extended mission \citep{Hirabayashi2021, Kikuchi2023}. 
The spacecraft will firstly flyby with (98943) Torifune (a.k.a. \CC) in 2026, and rendezvous with \KY in 2031. 

% 2. About 1998 KY26
\KY is an NEA discovered by the Spacewatch project in Arizona, US on 1998 May 28\footnote{\url{https://www.minorplanetcenter.net/mpec/J98/J98L02.html}} \citep{McMillan2007}.
Quick response observations of \KY soon after its discovery were performed in June 1998: radar observations using the Goldstone Solar System Radar and 
photometric observations with four small aperture ($D\leq 1$~m) telescopes \citep{Ostro1999}.
From the optical photometry, a rotation period of \KY was derived as $P=10.7015\pm0.0004$~minutes (1$\sigma$ error),
and color indices were derived as $B-R=0.083\pm0.070$, $V-R=0.058\pm0.055$, and $R-I=0.088\pm0.053$ (1$\sigma$ error with solar colors subtracted).
From the color indices as well as geometric albedo derived from their radar-derived diameter and absolute magnitude, they inferred that \KY belongs to B-, C-, F-, G-, D-, P-, or M-type asteroids in Tholen taxonomy \citep{Tholen1984}.
Since the radar albedo of \KY is not consistent with that of typical M-type asteroids, a classification as B-, C-, F-, G-, D-, or P-type was favored \citep{Ostro1999}.

The radar echo spectra, together with the derived rotation period, were used to estimate the \KY's shape and size.
Considering the uncertainties of its pole orientation, \cite{Ostro1999} estimated the effective diameter of \KY to be 30 ($\pm$ 10)~m.
The circular-polarization ratios (SC/OC ratios) of \KY were derived to be $0.5\pm0.1$, which exceeds 90\% of the SC/OC ratios of NEAs at the time of the publication.

% 3. Motivation
The Hayabusa2 extended mission gives us an exceptional opportunity to compare and test our knowledge of tiny asteroids obtained with telescopic and in situ observations. 
However, no detailed physical characterizations of \KY have ever been done except for colors, shape model, and rotation period \citep{Ostro1999}, simply due to the observational difficulties and poor observational opportunities.
As written by \citet{Ostro1999}, it is still unclear whether tiny asteroids are \textit{coherent solid rocks or gravitationally bound, multicomponent agglomerates}; \KY is a potential target to answer this question.
Moreover, \citet{Seligman2023} reported unexpected large out-of-plane nongravitational accelerations for seven NEAs including \KY.
One possibility is that some fractions of small NEAs are extinct "dark comets" in the near-Earth space.
Future characterizations are essential to examine the dynamical evolution of such a new population.
The determination of the diameter of \KY with an independent technique is therefore greatly desired 
to prepare and maximize the outcomes of the in situ measurements by the Hayabusa2 extended mission in 2031.

% 4. Outline of this paper
In this paper, we report the results of thermal infrared observations of \KY in May 2024.
The paper is organized as follows.
In Section 2, we describe our thermal infrared observations and data reduction with VLT/VISIR.
The results of the observations are presented in Section 3.
The constraints on the physical properties of \KY 
and the implications for the Hayabusa2 extended mission are discussed in Section 4.

\section{Observations and data reduction}\label{sec:obs}
An attempt to detect \KY at thermal infrared wavelengths was conducted in May 2024. 
As part of the ESO observing program 
``Thermal Observations of the Hayabusa2 Extended Mission Target Asteroid 1998 KY26"
(113.26MY, PI: J. Beniyama), we conducted measurements in the N-band
filters B8.7, B10.7, and B11.7 on three consecutive nights (2024 May 26, 27, and 28 UT). 
The observing conditions are summarized in Table \ref{tab:obs}.

We performed thermal infrared imaging observations using 
the VLT spectrometer and imager for the mid-infrared \citep[VISIR,][]{Lagage2004} 
located at the Cassegrain focus of Unit Telescope 2 of the ESO Very Large Telescope (VLT) on Cerro Paranal, Chile.
VISIR provides imaging at high-sensitivity in two mid-infrared atmospheric windows: 
the N-band between $\approx$ 8 to 13 µm and the Q-band between $\approx$ 16.5 to 24.5 µm.
VLT/VISIR reaches diffraction-limited images at $\sim0.3$~arcsec full width at half maximum (FWHM) in the N-band.

VISIR has a 1k $\times$ 1k Raytheon Aquarius IBC detector (1024 $\times$ 1024 pixel).
The field of view is 38.0~arcsec $\times$ 38.0~arcsec
with a pixel scale of 0.045~arcsec~pixel$^{-1}$.
The observations of all standard stars were done in the perpendicular 
nod-chop mode with a throw of 8~arcsec,
while the observations of \KY were done in the parallel nod-chop mode with a throw of 8~arcsec, which is recommended for faint sources.

The observing sequence with calibration stars interleaved with our science target covered more than 3 hours each night. 
The effective on-source integration times are between 15 and 18 minutes in each filter.
All measurements were taken in sky conditions with V-band seeing below or around 1.4~arcsec, precipitable water vapor (PWV) levels 
well below 5\,mm H$_2$O, and with the target seen under an airmass below 1.25.
Chopping frequencies were set to $\sim$4~Hz.
Although the Srehl ratio was in the range 0.4--0.5, the PSF FWHM was maintained to less than 0.4~arcsec.

% Observations of KY26
% Tool: https://www.eso.org/sci/php/phase2/ephemConverter/
\KY was observed in non-sidereal tracking 
with blind preset and offset speed from
VLT-centric \KY ephemeris 
provided by JPL Horizons\footnote{\url{https://ssd.jpl.nasa.gov/horizons}}, 
using observatory code 309 for Paranal.
% === This statement is about the orbit accuracy  (by Olivier) ===
Since \KY was recovered on 2024 April 17\footnote{\url{https://minorplanetcenter.net/mpec/K24/K24H23.html}} before our observations, 
its updated orbit was accurate at the $\leq1$~arcsec level during our observing runs.
The absolute pointing accuracy of the VLT is set by its pointing model, which is better than 3~arcsec RMS.
For all observations, it is further refined by the acquisition of a guide star, bringing the pointing accuracy to that of the star's position 
(typically better than 1~arcsec, even considering that its proper motion is not accounted for by the VLT). 
An additional source of pointing uncertainty is the time difference between the ephemerides time-stamp used and the time 
at which the telescope reaches that position and starts tracking at non-sidereal rates. 
While the goal is to keep that difference below a few seconds it can, in some cases, 
reach 10 or even 20~s. Accounting for the apparent motion of the object (800~arcsec/h at the time of the observations, or 0.22~arcsec/s),
this could cause a pointing error of up to 4--5~arcsec. 
Overall, we are therefore confident that the object was well within VISIR’s field of view. 
As tracking accuracy is exquisite (VLT performance webpage\footnote{\url{https://www.eso.org/sci/facilities/paranal/telescopes/ut/utperformance.html}}) 
and as the telescope was kept locked on its guide star during the observations, 
we are therefore confident that the object remained at a small ($<$4--5 arcsec), 
constant, offset from the expected position.
Co-adding the images blindly relying on the telescope position therefore does not degrade the image quality of the stack.
% === This statement is about the orbit accuracy  ===

\begin{deluxetable*}{rllc rccc ccccc}
\tabletypesize{\footnotesize}
\tablecaption{\label{t7}Summary of the observations\label{tab:obs}}
\tablehead{
    Starting Time & Object & Filter & $t_\mathrm{exp}$ & $N_\mathrm{img}$ & V     & $\alpha$ & $\Delta$ & $r_\mathrm{h}$ & Airmass & Sky & V seeing & MIR seeing \\
    (UTC)     &        &        & (s)              &                  & (mag) & (deg)    & (au)     & (au)           &        & &  (arcsec) & (arcsec) 
}
\startdata
2024-05-26 06:11:44& HD 145897 & B10.7 & 59.994 & 2 & 5.239 &  & &  & 1.11--1.11 & PH & 1.0 & 0.3\\
06:48:42& 1998 KY$_{26}$ & B10.7 & 101.772 & 20 & 20.2 & 41.4 &0.036 & 1.040 & 1.04--1.05 & PH & 1.1 & \\
07:44:41& HD 187150 & B8.7 & 59.994 & 2 & 6.46 &  & &  & 1.03--1.03 & PH & 1.0 & 0.3 \\
07:55:00& 1998 KY$_{26}$ & B8.7 & 103.95 & 20 & 20.2 & 41.5 &0.036 & 1.040 & 1.04--1.08 & PH & 1.2 &\\
08:50:52& 1998 KY$_{26}$ & B11.7 & 93.5712 & 20 & 20.2 & 41.6 &0.036 & 1.040 & 1.10--1.18 & PH & 1.0 &\\
09:44:11& HD 216149 & B11.7 & 60.1236 & 2 & 5.4 &  & &  & 1.10--1.10 & PH & 0.9 & 0.4\\
2024-05-27 05:12:33& HD 133774 & B11.7 & 60.1236 & 2 & 5.2 &  & &  & 1.11--1.11 & PH & 1.3 & 0.4 \\
05:21:58& 1998 KY$_{26}$ & B11.7 & 93.5712 & 20 & 20.2 & 44.4 &0.035 & 1.038 & 1.14--1.25 &  PH & 1.2 &\\
06:07:13& 1998 KY$_{26}$ & B8.7 & 103.95 & 20 & 20.2 & 44.5 &0.035 & 1.038 & 1.06--1.13 & PH & 1.3 &\\
07:11:10& HD 187150 & B8.7 & 59.994 & 2 & 6.46 &  & &  & 1.06--1.06 & PH & 1.3 & 0.3\\
07:14:41& HD 187150 & B10.7 & 59.994 & 2 & 6.46 &  & &  & 1.05--1.05 & PH & 1.3 & 0.3\\
07:20:28& 1998 KY$_{26}$ & B10.7 & 101.772 & 20 & 20.2 & 44.7 &0.035 & 1.038 & 1.04--1.04 & PH & 1.4 & \\
2024-05-28 06:24:13& HD 163376 & B10.7 & 59.994 & 2 & 4.863 &  & &  & 1.05--1.05 & CL & 0.7 & 0.3\\
06:41:17& 1998 KY$_{26}$ & B10.7 & 101.772 & 20 & 20.2 & 48.2 &0.033 & 1.035 & 1.05--1.10 & TN & 0.7 & \\
07:35:13& 1998 KY$_{26}$ & B11.7 & 93.5712 & 20 & 20.2 & 48.3 &0.033 & 1.035 & 1.04--1.05 & TN & 0.9 & \\
08:21:21& 1998 KY$_{26}$ & B8.7 & 103.95 & 20 & 20.2 & 48.4 &0.033 & 1.035 & 1.05--1.08 & CL & 0.8 & \\
09:21:07& HD 189831 & B11.7 & 60.1236 & 2 & 4.764 &  & &  & 1.05--1.05 & CL & 1.0 & 0.3\\
09:24:35& HD 189831 & B8.7 & 59.994 & 2 & 4.764 &  & &  & 1.06--1.06 & CL & 1.0 & 0.3\\
\enddata
\tablecomments{
    The observation time in UT at the starting-time of exposure (Starting Time), 
    target of observations (Object),
    filters (Filter),
    total exposure time ($t_\mathrm{exp}$), 
    number of images ($N_\mathrm{img}$),
    and V-band apparent magnitude (V) 
    are listed. 
    The phase angle ($\alpha$), 
    distance between \KY and observer ($\Delta$), 
    and distance between \KY and the Sun ($r_\mathrm{h}$) 
    at the observation starting time were from NASA JPL Horizons 
    on 2025 January 10.
    Elevations of \KY used to calculate the airmass range (Airmass) are also from NASA JPL Horizons.
    Sky condition (Sky, PH: Photometric, CL: Clear, TN: Thin cirrus) and
    seeing FWHM at 500~nm and zenith (V Seeing)
    are from the VLT UT2 Service Telescope Reports.
    Seeing FWHM of the standard observations (MIR seeing) are 
    derived by fitting light-intensity profiles with Moffat functions.
    }
\end{deluxetable*}

% Data reduction (by Eric)
The data were initially processed using a private and customized version of the VISIR pipeline \citep{EP_HDR}. For every day/filter dataset, the chopping and nodding offset frames are combined to produce a single frame in which the different "beams" are supposed to be. The observations were performed in the so-called "parallel mode" in which the signal is spread on the detector, with a central positive beam containing half of the total source flux and two negative beams symmetrically apart from the central beam \citep{EP_HDR}. At that stage, no point source is identifiable in the resulting frames. Yet, since the secondary mirror (chopper) and telescope nodding offsets are known from the telemetry, it is possible to blindly combine the beams by applying the adequate frames offsets and sum. This final image is radiometrically calibrated using the conversion factor from ADU to Jy deduced from the standard star observations obtained before or after \KY.

\section{Results} \label{sec:result}
\begin{figure*}[h!tb]
\resizebox{18cm}{!}{\includegraphics{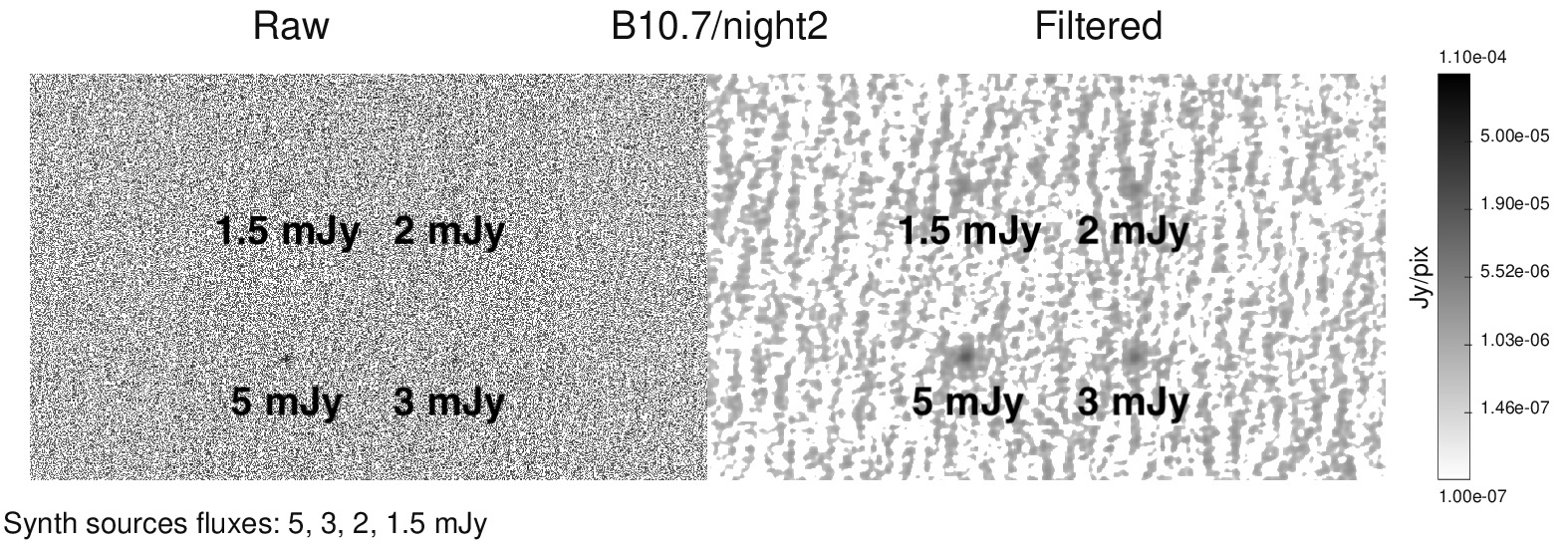}}
\caption{
    Final stacked image with synthetic point source objects with flux densities of
    5~mJy, 3~mJy, 2~mJy, and 1.5~mJy inserted.
    The flux density of the synthetic sources are indicated in the accompanying text.
}
\label{fig:nondetection}
\end{figure*}

We performed thermal observations with VLT/VISIR on three consecutive nights.
After the operations described in Section \ref{sec:obs}, one is searching for a single point source that could be \KY. 
Its position on the detector cannot be perfectly known since a blind telescope absolute preset is only accurate to a few arcsec. 
From the noise only, it is possible to assess a point source sensitivity. 
These sensitivities are given in Table \ref{tab:sensitivity}. 
None of the final reduced images in any filter/night is at first sight 
showing any presence of a point source that would correspond to \KY. 
In order to search deeper, the following procedure has been applied. 
For a given filter among the three selected for the observations, 
the best observation in terms of sensitivity is selected. 
It turned out that since the second night (2024 May 27) was the best one, 
all the selected observations come from this night. 
This image is then convolved with the center symmetric version of the PSF, 
which is equivalent to the correlation product with the PSF. 
This corresponds to the so-called optimal or "matched filtering". 
% (see \cite{W_MF}).
% I feel citing Wikipedia is not preferred. 
% EP : If the wikipedia page is well documented and correct, I see no problem. 
% It is more and more accepted. 
% There is no article dealing w/ matched filters. 
% Maybe textbooks but it is becoming more and more difficult to have access to them unfortunately.... 
% Last time I went to the university library, shelves very almost empty ..  -> I see.
This linear filtering allows to optimally increase the SNR and the chances of detection.
The corresponding point source detectivities, deduced from visual detection of the synthetic point sources with varying fluxes, are given in Table \ref{tab:sensitivity}. 
Despite this processing, no point source can be identified in any of the filtered datasets. 
In order to double-check visually the point sources detectivities 
deduced only from the noise measure, we also included synthetic point sources in the datasets. 
5, 3, 2, and 1.5~mJy point sources were artificially added to the frames (before matched filtering) using the measured PSFs. 
Figure \ref{fig:nondetection} displays the frames in the B10.7 most sensitive filter (2024 May 27) before filtering and after. 
Before filtering, the 1.5~mJy, 2~mJy, and 3~mJy sources are barely detectable; 
after filtering, the 2~mJy and 3~mJy sources become fairly detectable; which we consider as our detection upper limit.

\begin{deluxetable*}{rllc rccc cccc}
\tablecaption{
    Measured point source sensitivities in units of mJy/10$\sigma$ on the various datasets. 
    \label{tab:sensitivity}
    }
\tablehead{
    ~ & 2024 May 26 & 2024 May 27 & 2024 May 27 & 2024 May 28\\ 
    ~ & Raw & Raw & Filtered & Raw
    }
\startdata
B8.7 & 4.46  & 4.20 & 2.00 & 5.34  \\ 
B10.7 & 4.53 & 4.46 & 2.50 & 5.42  \\ 
B11.7 & 5.91 & 5.66 & 3.00 & 6.46 \\
\enddata
\tablecomments{
    "Raw" refers to the final image after pipeline processing, while "Filtered" refers to the same image after matched filtering (see the text).
    }
\end{deluxetable*}

\section{Discussion}
% 1 The fact: non-detection
Despite the good weather conditions, \KY was not detected.
The object's fast rotation and the three-nights repeated multi-band
observing sequences spanning much longer than the rotation period make it also unlikely that we saw each time only
the minimum cross-section of an elongated body.
The most-constraining non-detection limit from the VISIR measurement is
related to the B10.7 measurements from 2024 May 27; the optimum
filtered images would have shown \KY if the object had a flux density of 2~mJy or higher (see Section \ref{sec:result}).

% 2. TPM predictions for 1998~KY26
We calculated for this specific epoch, and the B10.7 reference wavelength at 10.64\,$\mu$m, a range of thermal model predictions to see if the non-detection
limit can tell us something about \KY's physical and thermal properties.
We started with the fast rotating model \citep[FRM,][and references therein]{Lebofsky1989} 
which has a temperature
distribution that is isothermal in longitude and depends on the latitude.
The use of the FRM in this case can be justified by the object's very short rotation period.
In parallel, we also used the near-Earth asteroid thermal model \citep[NEATM,][]{Harris1998}, 
using beaming parameters of $\eta$ = 1.0 and 1.5 
to account for the large phase angle ($\alpha$ = 45$^{\circ}$ on 2024 May 27). 
In addition, we used a
more sophisticated thermophysical model (TPM), which is based on the work by \cite{Lagerros1996, Lagerros1998, Muller2002}. 
The TPM allows to use of arbitrary spin-shape solutions, 
and considers a one-dimentional heat conduction into the surface, self-heating and
shadowing effects due to surface topography and roughness
for the true solar illumination and observing geometries as seen by VISIR.
Surface roughness is modelled by means of different parametrizations,
either as hemispherical segments or as random Gaussian surfaces.
The model parameters for the FRM, the NEATM, and the TPM are listed in Table \ref{tbl:model_input}.
It is known that a rotation period determined from only one apparition has some ambiguity  
especially if the shape is not very elongated as \KY \citep[e.g.,][]{Harris2014}.
We used the TPM with a rotation period of 5.35~minutes,
which is twice as fast as that reported in \cite{Ostro1999}.
It was determined by a recent study on lightcurves of \KY (T. Santana-Ros et al. 2025, in review).
We used a spherical shape in the TPM, which is justified with the 
long integration time and the short rotation period of \KY; we measured the rotational-averaged signal.

\begin{figure}[h!tb]
\resizebox{8cm}{!}{\includegraphics{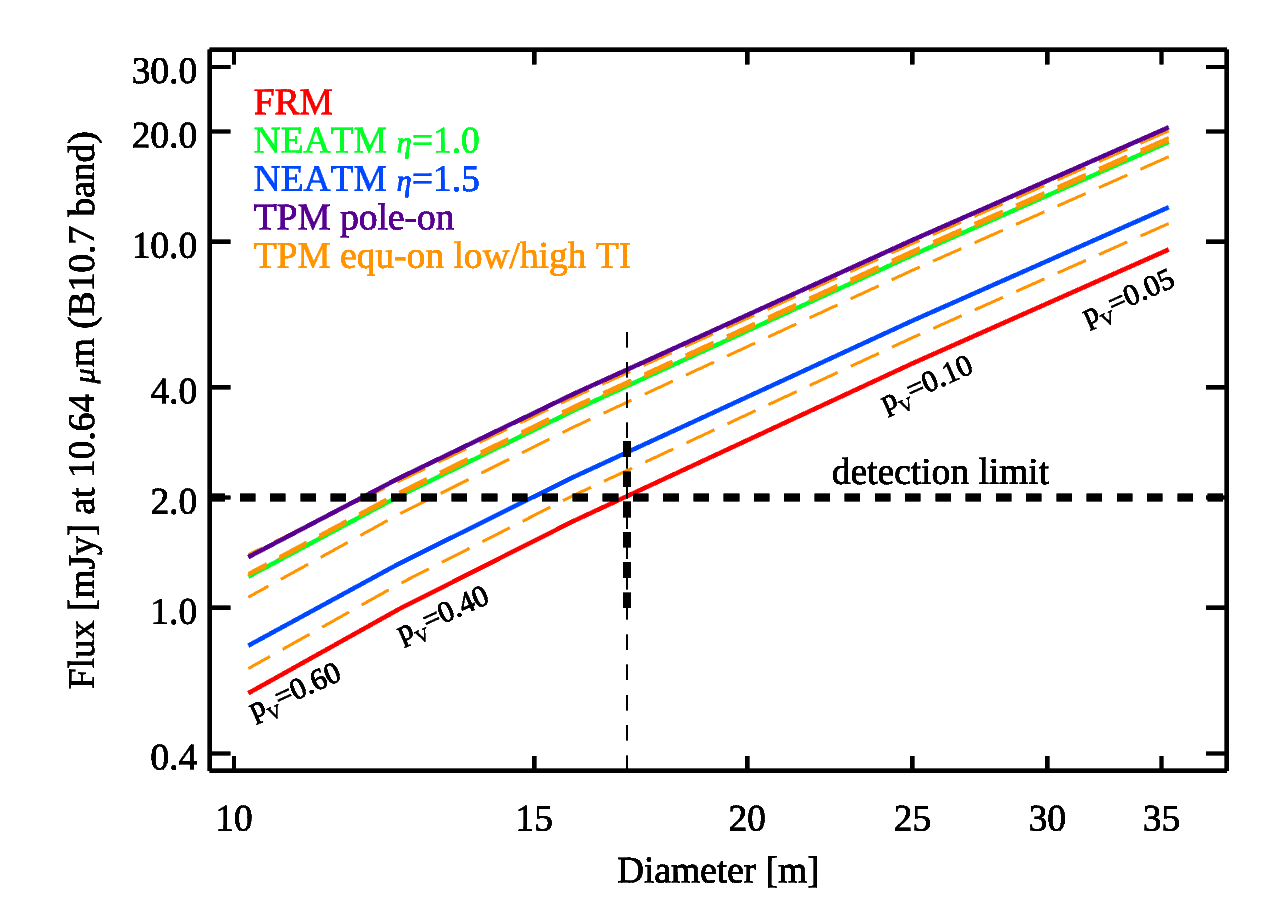}}
\centering
\caption{
    Flux predictions (in mJy for \KY for 2024 May 27 based on different model
    assumptions and for a wide range of object sizes. The VISIR/B10.7 band point-source
    detection limit is indicated as a horizontal dashed line. The various model 
    parameters are explained in the text.
    \label{fig:tpm_frm10}
    }
\end{figure}

% 3. Results of TPM
The highest fluxes are obtained in case of a pole-on viewing geometry and a low albedo
of $p_V$ = 0.05, where the thermal inertia have very little influence, see Figure \ref{fig:tpm_frm10}. 
The prograde equator-on geometry in combination with a low albedo and a low thermal inertia 
produced similar-level fluxes.
The retrograde equator-on geometry in combination with a high albedo ($p_V$=0.6) and a high thermal inertia 
(here $\Gamma$=300\,J\,m$^{-2}$\,s$^{-0.5}$\,K$^{-1}$)
gave the lowest fluxes. 
Higher thermal inertia, or a smooth surface (instead of a low-roughness surface) 
would lower the fluxes a few percent and push them into the regime of the FRM predictions. 
The FRM predictions, at this phase angle of about 45$^{\circ}$, 
produced
the lowest fluxes for a given object size. These values are considered as absolutely lowest flux
boundary, however, there is no evidence that small, fast-rotating asteroids are following the
isothermal temperature distributions of the FRM.
Recently, the thermal inertia of two fast-rotating small asteroids 
like \KY was reported as low as 10--100\,J\,m$^{-2}$\,s$^{
-0.5}$\,K$^{-1}$ \citep{Fenucci2021, Fenucci2023b}, 
though this is still not conclusive since the sample is limited.
The NEATM predictions are between the highest and lowest fluxes.

% 4. Constrains of diameter (and pV)
On the basis of Figure \ref{fig:tpm_frm10}, 
we conclude that \KY has to be smaller than 17~m otherwise
our VISIR measurements would have detected the asteroid. 
% Safely remove sentences below.
%   Considering only the more realistic TPM prediction, we can push the upper size limit to about 16\,m. 
%This value is compatible with the smaller end of the radar-derived 30 ($\pm$ 10)\,m by \cite{Ostro1999}.
This value is smaller than the radar-derived 30 ($\pm$ 10)\,m by \cite{Ostro1999}.
Our upper size limit of \KY is consistent with a recent study on lightcurves 
and a re-interpretation of the radar measurements (T. Santana-Ros et al. 2025, in review),
which indicates that \KY is truly smaller than the size estimated in \cite{Ostro1999}. 
The thermal infrared flux is driven by the size of the object and its surface temperature, which has a weak dependence on geometric albedo.
Thus, it is difficult to constrain geometric albedo with our non-detection.

% 5. Comparison with SST/IRAC non-detection
\cite{Mommert2014a} constrained physical and thermal properties of another tiny NEA 2009~BD ($H=28.4$) from
Spitzer Space Telescope/Infrared Array Camera (IRAC) non-detection at 4.5~$\mu$m.
They did not detect the target within 25~hr of integration.
They performed a TPM of 2009~BD as in our case.
Additionally, they model two nongravitational effects, solar radiation pressure and the Yarkovsky effect,
to constrain parameters such as size, thermal inertia, mass density, and obliquity (i.e., spin orientation). 
The combination of the two models gave stronger constraints on the physical properties of asteroids.
An unexpectedly large out-of-plane nongravitational acceleration was reported for \KY \citep[][which posed the possibility that \KY is a dark comet]{Seligman2023},
which add further complexity to the model.
% Hence, we are not confident that an analysis of the orbital 
% nongravitational acceleration using the Yarkovsky effect model could give meaningful information.

% 6. Recent studies
%   Kareta & Moskovitz 2024, DPS, D-type
%   Bourdelle de Micas+2024, EPSC, C- or B-type (not mention since the result is not written in the abstract)
\cite{Kareta2024a} reported multi-filter photometry of \KY in the visible wavelength with the 4.3~m Lowell Discovery Telescope.
They showed that \KY has a D-type reflectance spectrum, 
which might have primitive organic materials on the surface \citep{Barucci2018}.
A typical geometric albedo of D-types is estimated as 
$\sim0.06$ for asteroids with the diameter of $D\geq10$~km \citep{Usui2013},
while recently it was reported to be $\sim0.04$ for NEAs observed by the MIT-Hawaii NEO Spectroscopic Survey \citep{Marsset2022a}.
However, a caveat is that a typical albedo value for D-type NEAs with the diameter of $D\leq20$~m as \KY is unclear since the number of samples is limited.

On the other hand, it is possible to constrain the lower limit of 
the geometric albedo of \KY from our upper size limit and the updated 
absolute magnitude of $H=26.13\pm0.16$, 
which was derived from an extensive photometric study covering a wide range of solar phase angles and geometries
(T. Santana-Ros et al. 2025, in review).
The geometric albedo of \KY is estimated to be $\geq 0.19$ with $H=26.29$ 
and $D \leq 17$~m, which is significantly larger than the typical geometric albedo of D-types.
The large geometric albedo of \KY is consistent with a recent Xe-type classification in Bus-DeMeo taxonomy \citep{Bolin2025}.
Spectroscopic observations in wide wavelength coverage are essential to further investigate the surface composition of \KY.

% 7. Implications for Hayabusa2 extended mission
Our size constraint on \KY affects the operation of Hayabusa2 spacecraft.
A tight size constraint is obviously important during
proximity operations using remote sensing instruments since
pixel resolution and spatial resolution are essential in 
determination of the spacecraft's orbit.
Moreover, the Hayabusa2 spacecraft has a tantalum projectile on it \citep{Kikuchi2023}.
An impact experiment using the remaining projectile would be an option for the extended mission at \KY,
though a potential use of the projectile is under discussion \citep{Hirabayashi2021}.
Given a fixed bulk density and smaller size, an expected crater diameter might be larger.

\begin{deluxetable*}{ll}
\tablecaption{
    Thermal model parameter space for predicting
    KY26 fluxes for 2024 May 27 08:00 at the VISIR/B10.7 reference wavelengths of 10.64\,$\mu$m. 
    \label{tbl:model_input}
    }
\tablehead{
    Model & parameter space
    }
\startdata
FRM   & r = 1.038\,au, $\Delta$ = 0.0346\,au \\
\noalign{\smallskip}
NEATM & r = 1.038\,au, $\Delta$ = 0.0346\,au, $\alpha$ = 44.7$^{\circ}$  \\
      & beaming parameter $\eta$ = 1.0 and 1.5 \\
\noalign{\smallskip}
TPM   & r = 1.038\,au, $\Delta$ = 0.0346\,au, $\alpha$ = 44.7$^{\circ}$ \\
      & size (effective diameter) range: 10 ... 35\,m \\
      & geometric V-band albedo p$_V$: 0.05 ... 0.6 \\
      & H = 26.13, G = 0.15 \\
      & roughness (rms of surface slopes): 5 ... 20$^{\circ}$ \\
      & thermal inertia $\Gamma$: 30 ... 300\,J\,m$^{-2}$\,s$^{-0.5}$\,K$^{-1}$ \\
      & rotation period P: 5.35\, min \\
      & different spin-vector orientations: \\
      & (1) prograde, equator-on: \\
      & ($\lambda$, $\beta$) = (111.2$^{\circ}$, +77.1$^{\circ}$) \\
      & (2) retrograde, equator-on:\\
      & ($\lambda$, $\beta$) = (291.2$^{\circ}$, -77.1$^{\circ}$) \\
      & (3) pole-on:\\
      & ($\lambda$, $\beta$) = (291.2$^{\circ}$, +12.9$^{\circ}$) \\
\enddata
\tablecomments{
    $H$ is the latest value derived from an extensive photometric study (see the text).
    }
\end{deluxetable*}

\section{Conclusions}
We performed thermal observations of \KY, 
the Hayabusa2 extended mission rendezvous target, with VLT/VISIR in May 2024.
From the upper limit flux density via non-detection with VLT/VISIR,
we constrained that the diameter of \KY is smaller than 17~m through the TPM,
which is smaller than the radar-derived diameter.
Our size constraint on \KY is crucial for the operation of Hayabusa2 spacecraft
during proximity operations using remote sensing instruments
and a possible impact experiment using the remaining projectile on the spacecraft.

\begin{acknowledgments}
% Hayakawa fund? origin of this research (TBD)
We acknowledge the anonymous referee for a constructive report that improved the quality of the paper.
We thank Dr.~Bruno Leibundgut, the Director of the ESO for Science, for the time allocation to observe \KY. 
The authors are grateful to Dr.~Makoto Yoshikawa and Dr.~Toshi Hirabayashi for supporting this project. 
We thank Dr.~Mario van den Ancker for observing assistance with VLT/VISIR. 
J.B. thanks Dr.~Shigeyuki Sako and Dr.~Takafumi Kamizuka for the discussion of thermal observations. 
We are grateful to Dr.~Konrad R. W. Tristram, Dr.~Bin Yang, and Dr.~Julia V. Seidel for valuable comments on VLT/VISIR observations and data reduction.
This work is based on observations collected at the European Southern Observatory under ESO programme 113.26MY (PI: J. Beniyama)
This work was supported by JSPS KAKENHI Grant Numbers JP22K21344 and JP23KJ0640. 
Last but certainly not least, J.B. would like to thank support from the Hayakawa Satio Fund awarded by the Astronomical Society of Japan for giving the opportunity to work on this project.
\end{acknowledgments}

\facilities{VLT:Kueyen (VISIR)}
\software{
    NumPy \citep{Oliphant2015, Harris2020},
    pandas \citep{Reback2021},
    SciPy \citep{Virtanen2020},
    AstroPy \citep{Astropy2013, Astropy2018},
    astroquery \citep{Ginsburg2019}
}
\input {main.bbl}

\bibliographystyle{aasjournal}

\end{document}

%% file: main.bbl
\newcommand{\noopsort}[1]{} \newcommand{\singleletter}[1]{#1}